\documentclass[11pt,twoside]{article}
\usepackage{asp2006}
\usepackage{epsf}
\usepackage{graphics}
\usepackage{lscape}
\def\arg#1{{\it#1\/}}
\let\prog=\arg
\def\edcomment#1{\iffalse\marginpar{\raggedright\sl#1\/}\else\relax\fi}
\reversemarginpar

\begin{document}
\title{The Solar Microwave Flux and the Sunspot Number}
\author{L. Svalgaard  and H.~S. Hudson}
\affil{Space Sciences Laboratory, UC Berkeley}

\begin{abstract}
The solar F10.7 index is has been a reliable and sensitive activity index since 1947. As with other indices, it has been showing unusual behavior in the Cycle 23/24 minimum. 
The origins of the solar microwave flux lie in a variety of features, and in two main emission mechanisms: free-free and gyroresonance.
In past solar cycles F10.7 has correlated well with the sunspot number SSN. 
We find that this correlation has broken down in Cycle~23, confirming this with Japanese fixed-frequency radiometric microwave data.
\end{abstract}

\vspace{-0.5cm}
\section{Introduction}

The F10.7 daily solar index, introduced by Covington from 1947 (see Tapping et al., 2003, for more detail) has become a standard measure of solar activity and as a proxy for many variables not routinely measured.
\nocite{tapping03}
It represents the microwave flux density at 10.7~cm wavelength, which corresponds coincidentally to the electron Larmor frequency for $|${\bf B}$|$~=~10$^3$~G.
Shortly after its introduction, routine observations in Japan began at a set of four fixed frequencies (1.0, 2.0, 3.75, and 9.4~GHz), straddling the F10.7 frequency (2.8~GHz) and encompassing the gyroresonance signature in the solar microwave spectrum.
\citet{tanaka73} used these and other data to establish a uniform absolute photometric standard for the microwave range.

Microwave emission from the quiet Sun has many sources, all detectable via free-free emission or thermal gyroresonance radiation. 
Coronal holes may be bright, while filaments are dark.
A basic minimum level of emission somewhere below 70~SFU (one solar flux unit is 10$^{-22}$~W/m$^2$Hz) is presumed to come from the quiet background photosphere, and an ``S-component'' due to the other components, especially active regions and the enhanced network, varies on all time scales longer than those of flares.

\begin{figure}
\plotone{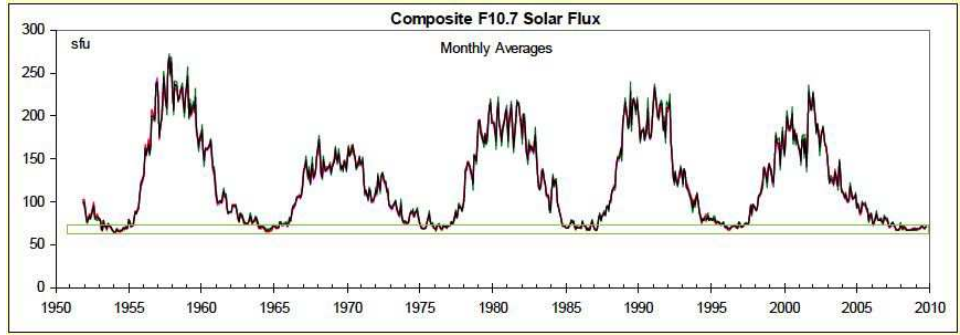}
\caption{\label{fig:ts}
The full length of the timeseries data for the microwave indices. 
We have scaled the four Nobeyama fixed frequefncies to F10.7 and overplotted them as monthly averages.
Note the stability of the minima (box).
}
\end{figure}

The Canadian (F10.7) and Japanese (four fixed frequency) measurements have continued in an uninterrupted time series up to the present.
The only substantial changes in the observing programs appear to have been the observing sites in each case: in 1991 the F10.7 facilities moved from Ottawa to Penticton, and in 1994 the Japanese facilities moved from Toyokawa to Nobeyama.
Each of these changes induced small but definite shifts of calibration, as we discuss further below.
The two microwave data sets compare extremely favorably with one another, as we show in Figure~\ref{fig:ts}.
An adjusted composite of the data shows that the six minima thus far observed all agree closely.

\section{Correlation with Sunspot Number}

\begin{figure}
\plotone{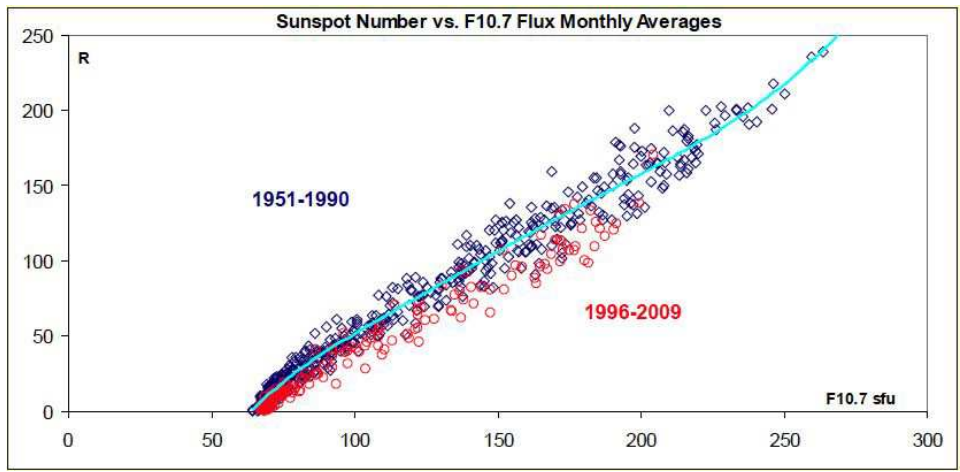}
\caption{\label{fig:corr}
Correlation of F10.7 monthly averages with sunspot number $R$ for two epochs: 1951-1988, and 1996-2009, with
empirical fits.
The correlation for cycle~23 systematically differs from the earlier pattern.
}
\end{figure}

The sunspot number is a much more qualitative index, but one that has an even longer history.
Since it reflects only sunspots, its solar origin is simpler than that of F10.7.
On the other hand it must depend upon observing conditions and observer bias.
Furthermore we really don't know how or why sunspots form.
The observed correlation between F10.7 and the sunspot number SSN is therefore not linear, but it was surprisingly precise and stable over 1951-1988 (Figure~\ref{fig:corr}).

\begin{figure}
\plotone{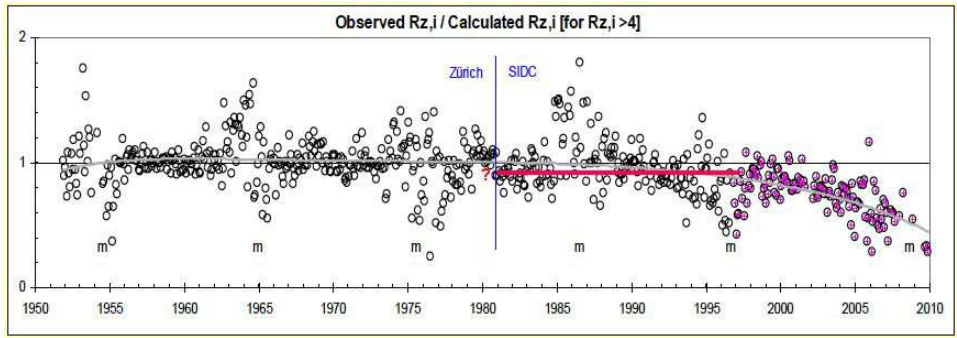}
\caption{\label{fig:norm}
Ratio of reported SSN to one predicted from the 1951-1988 interval of our composite microwave index.
The change in the institution responsible for the SSN in 1981 has a clear effect, but the relative diminution of F10.7 from the beginning of Cycle~23 is unmistakeable.
Here the ratio is shown only for times of SNN greater than 4.
The thick horizontal line shows the small calibration offset coincident with the SSN transfer from Z{\" u}rich to Brussels.
Small $m$s show the times of solar minima, during which the normalization has greater scatter.
}
\end{figure}

Following these first three maxima, as recorded in F10.7, the correlation changed. 
We show this in Figure~\ref{fig:corr} and use the 1951-1988 data to obtain a simple polynomial
fit.
This fit then allows us to estimate a value for the SSN, and we plot the ratio of the true SSN
to this predicted SNN in Figure~\ref{fig:norm}.
In this time series there are at least three risky epochs, in which major instrumentation changes occurred.
As mentioned, the F10.7 observations moved from Ottawa to Penticton in 1991 and the Japanese fixed-frequency data from Toyokawa to Nobeyama in 1994.
In addition the responsibility for SSN passed from Z{\" u}urich to Brussels (SIDC) in 1981.
Figure~\ref{fig:norm} does not correct for the systematic effect from the SSN change, which we estimate as a decrease of about 5\%, but we have adjusted for the microwave data changes in the composite F10.7 we use for the predicted SSN shown.
The relative decrease of SSN following the solar minimum of 1996 greatly exceeds any of the adjustments for the three risky epochs.

\section{Discussion}

These results make the case for a secular change in the relationship between sunspots and the sources of F10.7, which strongly reflects the active regions during times of elevated activity.
In principle this could reflect cycle-related changes of sunspot properties, as reported by Albregtsen et al. (1984) and by Penn and Livingston (1996).
Note that these results may differ, in that Livingston \& Penn (2009) suggest that this recent minimum may be qualitatively different: a multi-cycle trend may be appearing.
In this trend the brightness of sunspot umbrae apparently increases systematically, while the magnetic field decreases in a consistent manner.
These results were obtained with quite different techniques, in the case of Livingston \& Penn via quite simple Zeeman measurements  of an Fe~{\sc i} line in the infrared at 1.5648$\mu$.
These measurements also permit a clean differential measurement of sunspot umbral intensity relative to the neighboring quiet photosphere.

We do not claim that the F10.7 anomaly we report necessarily confirms the remarkable conclusions drawn by Livingston \& Penn, but the results are consistent with the idea that active-region magnetism systematically is growing less capable of sunspot formation.
This of course would be unprecedented and unexplained, but our records of solar activity only cover a few repeats of the notoriously variable solar cycle, with modern data.

\section{Conclusions}

We confirm the anomalous long-term behavior of the non-flare solar microwave fluxes as represented in the F10.7 index, reported elsewhere by W.~D. Pesnell and by K.~Tapping (personal communications, 2009).
Cycle 23 appears to have been anomalous when referred to the sunspot number SSN. 
We suggest that the anomaly really lies in a secular variation in the nature of sunspots, rather than in the active-region magnetism responsible for the microwave S-component as such.

\bigskip\noindent
{\bf Acknowledgements:}

{}

\end{document}